\documentclass{article}
\usepackage{a4wide}
\usepackage{amsmath}
\usepackage{epsfig}
\newcommand{\sech}{\operatorname{sech}}
\newcommand{\eqind}{\hspace{15mm}}
\title{Comment on `Replica analysis of the $p$-spin interaction Ising
spin-glass model'}
\author{Peter Gillin and David Sherrington\\
\footnotesize{Physics Department, University of Oxford, Theoretical
Physics, 1 Keble Road, Oxford OX1 3NP, UK}}
\begin{document}
\maketitle
\begin{abstract}
\noindent
We demonstrate that the analytic calculation of the 1RSB break point
parameter in a paper by de Oliveira and Fontanari\cite{of:ipsgh} is
erroneous, due to the omission of a higher order term in a lengthy
perturbative calculation, and provide a refinement of the accompanying
numerical results.

\end{abstract}

In 1999, de Oliveira and Fontanari (OF) studied the one-step replica
symmetry breaking (1RSB) of a glass of Ising spins with a quenched
random $p$-spin interaction of infinite range in a
field\cite{of:ipsgh}. The Hamiltonian is given by
\begin{gather}
\mathcal{H} = \sum_{i_1 < i_2 \dots < i_p} J_{i_1 \dots i_p}
\sigma_{i_1} \dots \sigma_{i_p} - h \sum_i \sigma_i
\end{gather}
where the $J_{i_1 \dots i_p}$ are independent Gaussian variables with
zero mean and variance $p!J^2/2N^{p-1}$. They found that for fields
$h$ less than a critical value $h_\mathrm{c}$ the transition was
discontinuous (D1RSB), while for $h>h_\mathrm{c}$ it is continuous
(C1RSB). In \S3.1 of that paper, they give certain results on the
C1RSB line, including an expression for the 1RSB break point quantity
$x$ ($m$ in their notation). We demonstrate an error in this
calculation of $x$. Moreover, they present numerical results within
the 1RSB phase which are (for reasons we shall explain below)
inaccurate near the C1RSB line. We present a discussion and refined
results.

The self-consistent equation for the RS phase is\cite{g:ipg2sg}
\begin{gather}
q = T(2)
\end{gather}
where
\begin{gather}
T(n) \doteq \int \! \frac{dz}{\sqrt{2\pi}}\, e^{-z^2/2} \tanh^n \left(
\sqrt{\tfrac{1}{2} p \beta^2 J^2 q^{p-1}}\, z + \beta h \right).
\label{scrs}
\end{gather}
This solution becomes unstable against small replica symmetry breaking
fluctuations on the Almeida--Thouless line, given by\cite{g:ipg2sg}
\begin{gather}
k S(4) = 1
\label{atrs}
\end{gather}
where
\begin{align}
k &\doteq \frac{1}{2} p(p-1) \beta^2 J^2 q^{p-2},
\\
S(n) &\doteq \int \! \frac{dz}{\sqrt{2\pi}}\, e^{-z^2/2} \sech^n \left(
\sqrt{\tfrac{1}{2} p \beta^2 J^2 q^{p-1}}\, z + \beta h \right).
\end{align}

The self-consistent equations for the 1RSB phase are\cite{of:ipsgh}
\begin{subequations}
\label{scrsb}
\begin{align}
q_0 &- \int \frac{dz_0}{\sqrt{2\pi}}\, e^{-z_0^2/2} \left( \frac{ \int
\frac{dz_1}{\sqrt{2\pi}}\, e^{-z_1^2/2} \cosh^x G \, \tanh
G} { \int \frac{dz_1}{\sqrt{2\pi}}\, e^{-z_1^2/2} \cosh^x
G} \right)^2 = 0
\label{scrsb0}
\\
q_1 &- \int \frac{dz_0}{\sqrt{2\pi}}\, e^{-z_0^2/2} \; \frac{ \int
\frac{dz_1}{\sqrt{2\pi}}\, e^{-z_1^2/2} \cosh^x G \, \tanh^2
G} { \int \frac{dz_1}{\sqrt{2\pi}}\, e^{-z_1^2/2} \cosh^x
G} = 0
\label{scrsb1}
\\
\frac{1}{4} (p-1) \beta^2 J^2 (q_1^p - q_0^p) &= - \frac{1}{x^2} \int
\frac{dz_0}{\sqrt{2\pi}}\, e^{-z_0^2/2} \, \ln \int
\frac{dz_1}{\sqrt{2\pi}}\, e^{-z_1^2/2} \cosh^x G \notag \\& \eqind
+ \frac{1}{x} \int \frac{dz_0}{\sqrt{2\pi}}\, e^{-z_0^2/2}
\; \frac{ \int \frac{dz_1}{\sqrt{2\pi}}\, e^{-z_1^2/2} \cosh^x
G \, \ln \cosh G} { \int \frac{dz_1}{\sqrt{2\pi}}\,
e^{-z_1^2/2} \cosh^x G} = 0
\label{scrsbx}
\end{align}
\end{subequations}
where
\begin{gather}
G \doteq \sqrt{\tfrac{1}{2} p \beta^2 J^2 q_0^{p-1}} \, z_0 +
\sqrt{\tfrac{1}{2} p \beta^2 J^2 (q_1^{p-1} - q_0^{p-1})} \, z_1 +
\beta h\,.
\end{gather}
The C1RSB line is defined by $q_0=q_1$. In this case, $G$ does not
depend on $z_1$, and the integrals over this variable are trivial. We
observe that \eqref{scrsb0} and \eqref{scrsb1} reduce to the same
equation, namely \eqref{scrs}: on this line, the solution coincides
with the RS, as we would expect. We obtain a second piece of
information by subtracting these equations and performing a series
expansion in the quantity $\epsilon = q_1 - q_0$: \eqref{scrsb0} and
\eqref{scrsb1} both become
\begin{gather}
\label{ord1}
q = T(2) + O(\epsilon)
\end{gather}
and the difference $\eqref{scrsb1}-\eqref{scrsb0}$ becomes
\begin{gather}
\label{ord2}
\epsilon = k S(4) \epsilon + O(\epsilon)^2.
\end{gather}
So \eqref{atrs} is also satisfied on the C1RSB line, that is the
transition coincides with the onset of instability in the RS solution,
again as we would expect.

On the C1RSB line, \eqref{scrsbx} is trivially solved. We obtain
further information by a series expansion. To first order we get
\begin{gather}
\frac{1}{2} k q \epsilon = \frac{1}{2} k T(2) \epsilon +
O(\epsilon)^2.
\end{gather}
This simply tells us that \eqref{scrs} holds, which we already
knew. We therefore eliminate this first order term by subtracting
\eqref{scrsb0} multiplied by $k \epsilon / 2$ from \eqref{scrsbx} to
obtain a new equation. To second order we get, rearranging slightly,
\begin{gather}
\frac{1}{4} k \left( \frac{p-2}{q} [q-T(2)] + [1-kS(4)] \right)
\epsilon^2 + O(\epsilon)^3 = 0.
\end{gather}
This tells us that \eqref{atrs} holds, which again we already knew. We
therefore eliminate these second order terms by subtracting
\eqref{scrsb0} multiplied by $(p-2)k \epsilon^2/4q$ and
$[\eqref{scrsb1}-\eqref{scrsb0}]$ multiplied by $k \epsilon/4$ to
obtain another new equation. To third order we get
\begin{gather}
\label{ord3}
\frac{k^2}{24q} \left[ \frac{p-2}{k} C + 4qk S(4) - 6qk S(6) + 2qk
S(6) (1-x) \right] \epsilon^3 + O(\epsilon)^4 = 0
\end{gather}
where
\begin{gather}
C \doteq 2(p-1) - 2(p-3)\frac{T(2)}{q} - 3kS(4).
\end{gather}
We solve this to obtain an expression for $x$ near the C1RSB line:
\begin{align}
1-x &= \frac{ 6qk S(6) - 4qk S(4) - (p-2) C k^{-1} } {2kq S(6)} +
O(\epsilon)
\\
\label{xc}
&= \frac{ 6qk S(6) - [4qk + (p-2)] S(4) } {2kq S(6)} + O(\epsilon)
\end{align}
where we have used \eqref{ord1} and \eqref{ord2} to simplify our
expression. This differs from (36) and surrounding equations of OF. We
note that if one erroneously neglects the $O(\epsilon)$ terms of
\eqref{scrsb0} when subtracting that equation multiplied by $(p-2)k
\epsilon^2/4q$ in the above process (using only the leading order
equation \eqref{scrs} instead) one obtains an incorrect form of the
$O(\epsilon)^3$ equation \eqref{ord3} which gives exactly the form of
OF.

Since we have shown the analytic expression of OF for $x$ near the
C1RSB line to be incorrect, we must question the accuracy of their
numerical solutions of the 1RSB equations \eqref{scrsb} in that
region, as the latter appeared to corroborate the former. The
determination of $x$ is indeed rather delicate, as the $x$-dependence
of these equations is very weak, for reasons that are clear from the
above analysis: it appears in a term $O(q_1-q_0)^2$ smaller than the
leading order, and close to C1RSB, $(q_1-q_0) \ll 1$ by definition.

We adopt an approach designed to avoid this problem. Rather than
solving the equations as given, we choose a judicious linear
combination which does not possess the same flatness. We know from
above that subtracting \eqref{scrsb0} multiplied by $k (q_1-q_0) / 2$
from \eqref{scrsbx} eliminates the leading order of the latter,
leaving an equation where the $x$-dependence is suppressed only by a
factor $O(q_1-q_0)$; and that further subtracting \eqref{scrsb0}
multiplied by $(p-2)k(q_1-q_0)^2/4q$ and
$[\eqref{scrsb1}-\eqref{scrsb0}]$ multiplied by $k(q_1-q_0)/4$
eliminates the next order, leaving an equation whose leading order is
linear in $x$. We find it most efficient to use the second of these
very close to C1RSB (where the problem is worst) and the first
elsewhere. We use a modified form of Newton--Raphson to find the
roots, and do the numerical integration using Gauss--Hermite
quadrature.

The figure shows the solution at $p=3$ and $h/J=1$ as a function of
temperature. This is equivalent to the solid line in figure 3 of
OF. The main line shows the numerical solution for $x$. The two
diamonds show the predictions for $x$ on the C1RSB line, the lower
using \eqref{xc} and the upper using the equivalent expression of
OF. The inset shows an enlargement of the region around the
transition, with rectangles for the numerical predictions for $x$ and
a diamond for our perturbative result.

\begin{figure}
\begin{center}
\epsfig{file=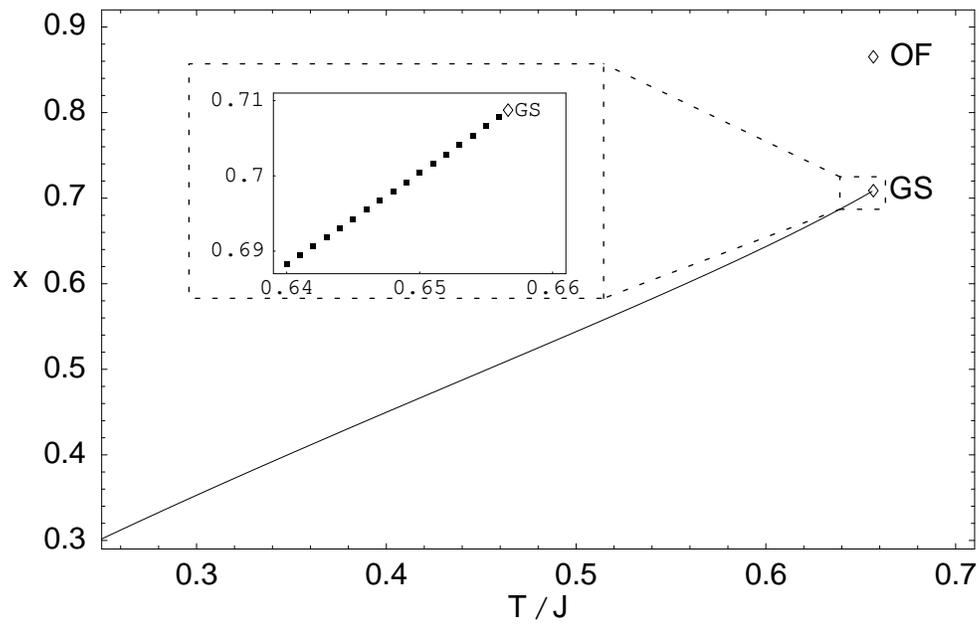,width=5.5in}
\caption{The numerical solution of \eqref{scrsb} for $x$ with $p=3$
and $h/J=1$, and a comparison with the analytic results at the
transition.}
\end{center}
\end{figure}

\section*{Acknowledgments}

We would like to thank EPSRC (UK) for financial support, PG for
research studentship 97304251 and DS for research grant
GR/M04426. Additionally, we thank Hidetoshi Nishimori and Jos\'e
Fernando Fontanari for useful discussions.


\begin{thebibliography}{1}

\bibitem{of:ipsgh}
V.~M. de~Oliveira and J.~F. Fontanari,
\newblock J. Phys. A {\bf 32}, 2285 (1998).

\bibitem{g:ipg2sg}
E.~Gardner,
\newblock Nucl. Phys. B {\bf 257}, 747 (1985).

\end{thebibliography}
\end{document}